\documentclass{aa}

\usepackage{graphicx}
\usepackage{txfonts}
\usepackage{psfrag}

\begin{document}

\title{Detection of the high $z$ \object{GRB 080913} and its implications on progenitors and energy extraction mechanisms. 
\thanks
{Based on observations taken with the 3.5m Calar Alto telescope, the Plateau de Bure interferometer and the ${\it XMM-Newton}$ satellite.} 
} 

\author{D.~       P\'erez-Ram\'{\i}rez\inst{1,2}
   \and A.~       de Ugarte Postigo  \inst{3}
   \and J.~       Gorosabel          \inst{4}
   \and M. A.~    Aloy               \inst{5}
   \and G.        J\'ohannesson      \inst{6}
   \and M. A.~    Guerrero           \inst{4}
   \and J. P.~    Osborne            \inst{2}
   \and K.~L.~    Page               \inst{2}
   \and R. S.~    Warwick            \inst{2}
   \and I.~       Horv\'ath          \inst{7}
   \and P.~       Veres              \inst{7,8} 
   \and M.~       Jel\'{\i}nek       \inst{4} 
   \and P.~       Kub\'anek          \inst{4,9}
   \and S.~       Guziy              \inst{4,10}
   \and M.~       Bremer             \inst{11}
   \and J.~M.~    Winters            \inst{11} 
   \and A.        Riva               \inst{12} 
   \and A. J.~    Castro-Tirado      \inst{4} 
      }

\offprints{D. P\'erez-Ram\'{\i}rez, \email{dperez@ujaen.es} } 
 
\institute{Departamento de F\'\i sica, Universidad de Ja\'en, 
           Campus Las Lagunillas, E-23071 Ja\'en, Spain.
       \and Department of Physics and Astronomy, The University of Leicester, 
           Leicester, LE1 7RH, UK.
       \and European Southern Observatory, Casilla 19001, Santiago 19, Chile.
       \and Instituto de Astrof\'\i sica de Andaluc\'\i a (IAA-CSIC), P.O. Box 3.004, E-18.080 Granada, Spain. 
       \and Departamento de Astronom\'\i a y Astrof\'\i sica, Universidad de 
            Valencia, C/ Dr. Moliner s/n, 
            E-46100 Burjassot, Valencia, Spain. 
       \and Department of Physics and SLAC National Accelerator Laboratory, Stanford University, Stanford, CA 94305, USA.
       \and Dept. of Physics, Bolyai Military University, POB 15, 1581 Budapest, Hungary.
       \and Dept. of Physics of Complex Systems, E\"otv\"os University, P\'azm\'any P. s. 1/A, 1117 Budapest, Hungary.
       \and Edif. Institutos de Investigaci\'on (GACE-ICMOL), Universidad de Valencia, Campus de Paterna, E-46980 Paterna, Valencia, Spain.
       \and Nikolaev State University, Nikolskaya 24, 54030 Nikolaev, Ukraine. 
       \and Institute de Radioastronomie Milim\'etrique (IRAM), 300 rue de la 
            Piscine, 38406 Saint Martin d\' \rm H\'eres, France.
       \and INAF - Osservatorio Astronomico di Brera, Via Emilio Bianchi, 46, 
            Merate, LC, 23807, Italy.}

\date{Received / Accepted } 
 
\abstract {} {We present multiwavelength observations of one of the
  most distant gamma-ray bursts detected so far, GRB\,080913. Based on
  these observations, we consider whether it could be classified as a
  short-duration GRB and discuss the implications for the progenitor
  nature and energy extraction mechanisms.}  {Multiwavelength X-ray,
  near IR and millimetre observations were made between 20.7 hours and
  $\sim$16.8 days after the event.}  {Whereas a very faint afterglow
  was seen at the 3.5m CAHA telescope in the nIR, the X-ray afterglow
  was clearly detected in both ${\it Swift}$ and ${\it XMM-Newton}$
  observations.  An upper limit is reported in the mm range.  We have
  modeled the data assuming a collimated $\theta_0$ $\gtrsim$
  3$^\circ$ blast wave with an energy injection at $\sim 0.5$ days
  carrying $5\sim 10^{52}$ erg or approximately 12 times the initial
  energy of the blast wave.  We find that GRB\,080913 shares many of
  the gamma-ray diagnostics with the more recent burst GRB 090423 for
  being classified as short had they ocurred at low redshift. If the
  progenitor were a compact binary merger, it is likely composed by a
  NS and BH. The Blandford-Znajek (BZ) mechanism is the preferred one
  to extract energy from the central, maximally-rotating BH. Both the
  magnetic field close to the event horizon ($B$) and the BH mass
  ($M_{bh}$) are restricted within a relatively narrow range, such
  that $(B / 3\times 10^{16}\,\rm{G}) (M_{bh} / 7\,M_\odot ) \sim 1$.
  Similar constraints on the central BH hold for collapsar-like
  progenitor systems if the BZ-mechanism works for the system at
  hand.}  {}
 
          
\authorrunning{P\'erez-Ram\'irez et al.} 

\titlerunning{GRB 080913: an intrinsically short, hard and high $z$ GRB?} 

\maketitle 
 
\section{Introduction} 
 
Gamma-ray bursts (GRBs) are generally classified into two main groups
(Kouveliotou et al. 1993), those with short duration and hard spectra
and those with long duration and soft spectra. This simplistic
classification scheme could be more complex, as shown by several
studies (Zhang et al. 2009; Horv\'ath et al. 2006, 2008).

For short bursts the general idea is that they originate in the near
Universe, at redshifts significantly lower ($z$$\sim$0.5) than
those of long GRBs (e.g. GRB 060502b at $z$ = 0.287, Bloom et
al. 2007; GRB 051221a at $z$ = 0.5464, Soderberg et al. 2006; GRB
050911 at $z$ = 0.1646, Berger et al. 2007a; GRB 050724 at $z$ =
0.257, Berger et al. 2005; GRB 050709 at $z$ = 0.160, Fox et al. 2005;
GRB 050509b at $z$ = 0.226, Gehrels et al. 2005).  However, de Ugarte
Postigo et al. (2006) observed GRB\,060121 (T$_{90}$=2s; Arimoto et
al. 2006) and provided a most probable photometric redshift of $z$ =
4.6. They suggested that this burst could be the first of a class of
short gamma-ray bursts residing at high redshift, which probably
belongs to a different progenitor group. Short high-redshift bursts
were later studied statistically by Berger et al. (2007b), who found
that a significant number of distant short bursts could exist.

GRB 080913 was discovered by {\it Swift} on 13 Sep 2008 (Schady et al.
2008). The burst started at 06:46:54 UT and lasted for $\approx8$~s,
placing it, at first sight, in the \textit{long-duration} class of
GRBs (Stamatikos et al. 2008).  It was also observed by {\it
  Konus}/WIND and had a fluence of (5.6 $\pm$ 0.6) $\times$ 10$^{-7}$
erg cm$^{-2}$ in the 15-150 keV range, making it an average GRB. In
the combined BAT-WIND spectrum, the observed prompt energy spectrum
could be best-fitted by a power-law with an exponential cutoff model
$d$N/$d$E $\sim$ E$^{\alpha}$ $\times$ exp(-(2+{$\alpha$}) $\times$
E/E$_{peak}$) with $\alpha$ = $-0.89_{-0.46}^{+0.65}$ and E$_{peak}$ =
131$_{-48}^{+225}$ keV (Pal\' \rm shin et al. 2008).

The prompt dissemination (21 s) of the GRB position by {\it Swift}
enabled instant responses of robotic telescopes, such as the REM
robotic telescope (D\' \rm Avanzo et al. 2008). Rapid observations
obtained by GROND at the 2.2m telescope in La Silla allowed the identification
of a near-infrared (nIR) counterpart (Rossi et al. 2008)
$\sim$ 3 min after the burst trigger.  Shortly after {\it Swift}
slewed and started data acquisition, a fading X-ray source was
detected by the {\it Swift}/XRT, which was identified as the GRB
080913 afterglow (Beardmore et al.  2008). This triggered a
multiwavelength campaign at different observatories aimed at studying
the afterglow. A photometric redshift in the range 6.1-6.7 was derived
(Greiner, Kruehler and Rossi 2008), and a spectroscopic $z$ = 6.7 was
later confirmed by a VLT spectrum (Fynbo et al. 2008, Greiner et
al. 2009).  This implied rest frame values of T$_{90}$ $\sim$ 1 s and
E$_{peak,rest}$ $\sim$1000 keV consistent with a
\textit{short-duration} GRB (Pal\' \rm shin et al.  2008). This is
also supported by the negligible spectral time lags found in the BAT
energy range (Xu 2008).  For a standard cosmology model with H$_0$ =
71 km/s/Mpc, $\Omega_{\rm M}$ = 0.27, $\Omega_{\Lambda}$ = 0.73, the
isotropic energy release is E$_{iso}$ $\sim$ 7 $\times$ 10$^{52}$ erg
(1 keV$-$10 MeV, rest frame), with a look-back time of 13.67 Gyr.

A more recently detected burst, GRB\,090423 (Tanvir et al. 2009,
Salvaterra et al. 2009), is a further extreme redshift ($z$=8.2) and
potentially short GRB which exhibits similar properties, such as burst
duration (T$_{90}$ $\sim$ 10.3 $\pm$ 1.1s, and a rest-frame duration
of $\sim$ 1s), spectral lag times negligibly small, consistent with
zero, and an intrisincally hard spectrum as GRB\,080913.

This burst, together with GRB\,080913, points to the fact that the
current dichotomy is not always consistent. That is, standard
indicators of the physical nature of GRBs, such as duration and
hardness, may no longer be the only diagnostic used in physically
classifying high-redshift GRBs. They reveal a need for a revision of
the traditional observational criteria. Zhang et al. (2009) tackled
this question proposing new operational procedures in the
determination of the physical category of GRBs.  According to this
work, GRB\,080913 and GRB\,090423 are considered to belong to the Type
II category (i.e. massive-star core collapse origin). However, these
two high-$z$ bursts may also be compatible with a ''specific Type I
scenario'' driven by the Blanford-Znajek mechanism in a BH-NS merger.

Here we report multiwavelength observations carried out, from the
millimetre to the X-ray band, in order to study the afterglow of GRB
080913. We also discuss the implications of these observations for the
nature of short-duration GRBs progenitors. Finally, we include some
results extracted from the literature about the implications for the
nature of long-duration GRBs progenitors

\begin{figure} 
\begin{center}
      \resizebox{8.5cm}{!}{\includegraphics{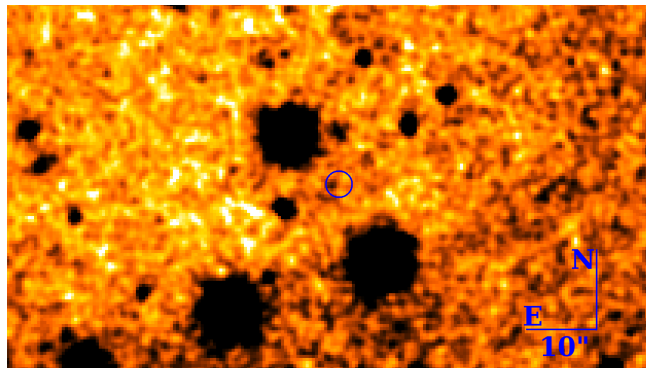}} 
      \caption{The $J$-band image of the \object{GRB 080913} field
               taken at the 3.5m CAHA telescope on 14 Sep 2008 (20.7
               hr after the burst onset).  The error circle marks the
               XRT position of the counterpart with a radius of
               1.9$^{\prime\prime}$ (Beardmore et al. 2008).}
    \label{carta ID}
\end{center} 
\end{figure}

\section{Observations and data reduction} 
  \label{observaciones}

  \subsection{Near-IR observations}   

Target of opportunity (ToO) observations in the nIR were triggered starting
20.7~hr after the event with the 3.5~m telescope (+OMEGA~2000; Bailer-Jones et
al. 2000) at the German-Spanish Calar Alto Observatory (CAHA). A 5,500 sec
image was acquired in the $J$-band filter (see Fig. 1), with a
1.35$^{\prime\prime}$ average seeing.  We followed the standard data reduction
procedures such as dark and sky frame subtraction, plus bad-pixel mask and
master flat-field correction.  The photometry for our final image was
performed by means of the PHOT routine under IRAF.\footnote{IRAF is
distributed by the NOAO, which are operated by USRA, under cooperative
agreement with the US NSF.}  A range of apertures were checked, and the one
yielding the minimum photometric error was selected. The candidate initially
reported by Rossi et al. (2008) was barely detected ($2.3 \sigma$ level) in
the $J$-band image (Fig. 1), with an estimated J-band Vega magnitude of
22.4$\pm0.5$, including the calibration zero point error (0.15 mag) given by
the 2MASS Catalogue.  In order to compare our detection with the lightcurve
for this GRB as presented by Greiner et al. (2009), we have evaluated the
AB-to-Vega system magnitude offset coefficient for the OMEGA~2000 instrument
in the $J$-band to be 0.97.  The converted J-band magnitude in the AB system
for our detection is 23.4$\pm0.5$, which at the time of the observation,
agrees with the GROND data (Greiner et al. 2009). Our detection occurred close
to the peak of the re-brightening phase (Fig.~\ref{lightcurve}).

  \subsection{Millimetre observations}

Additional mm observations were obtained at the Plateau de Bure
Interferometer (PdBI) as part of our ToO programme.  The PdBI observed
the source on different occasions in the period of time of three days
in compact configuration. We used the carbon star MWC349 as primary
flux calibrator (assuming $F(\nu)=1.1 \cdot (\nu/86.2 \rm GHz)^{0.6}$)
with the amplitude and phase calibrations relative to the quasar
0454-234.  The data reduction was done with the CLIC and MAPPING
software distributed by the Grenoble GILDAS group.  We analysed the
data with position-fixed (RA(2000) = 04:22:54.66, Dec(2000) =
-25:07:46.2) fits in the UV plane, which only yielded upper limits but
allows to constrain the mm-lightcurve. The 3$\sigma$-limits are 0.72
mJy (99 GHz, Sep 16.1 UT), 1.44 Jy (84 GHz, Sep 21.2 UT) and 0.90 mJy
(106 GHz, Sep 30.1 UT).

\begin{figure}[t] 
\begin{center}
      \includegraphics[width=6cm]{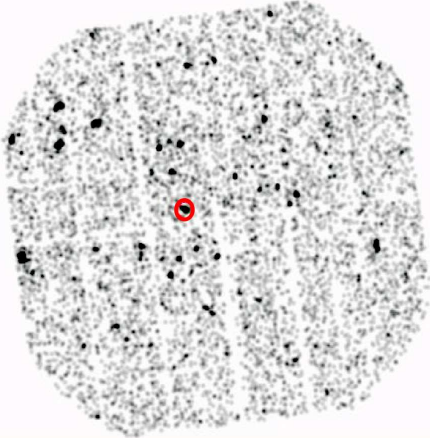}
      \caption{The EPIC-pn image of the high redshift GRB 080913
        obtained with {\it XMM-Newton} $\sim$4.5 days after the burst
        onset.  The image has been adaptively smoothed to emphasize
        the detection of the GRB. The position derived by {\it Swift}
        for this burst is marked by the circle.}
    \label{EPIC}
\end{center} 
\end{figure}

  \subsection{X-ray observations}
   
We made use of public {\it Swift}/XRT data obtained starting $\sim$94s
after the BAT trigger time (T$_{\rm 0}$).  The data were acquired in
the XRT Photon Counting (PC) mode.

The lightcurve in the 0.3-10 keV energy range (Evans et al. 2007) exhibits a
number of flares in the initial orbit, with the peak of the largest flare
observed at T$_{\rm 0}$+1800 s (in the observer frame).  The decay in the time
interval T$_{\rm 0}$+400 s to T$_{\rm 0}$+1100 s can be fitted by a power-law
with an estimated decay index $\alpha_X$ = 1.2$^{+0.2}_{-0.1}$ (where $f_X
\propto t^{-\alpha_X}$).  The spectrum corresponding to the period of flare
activity in the initial orbit, can be fitted by an absorbed power-law with a
photon index of $\Gamma$ $\sim$ 1.7$^{+0.5}_{-0.4}$, assuming that the
Galactic column density value in the direction of the burst is 3.2 $\times$
10$^{20}$ cm$^{-2}$ (Kalberla et al. 2005).
  
A ToO \emph{XMM-Newton} observation (Obs. ID.~0560191701) started on
Sep 17.61 UT, i.e. 4.3 days after the {\it Swift}/BAT trigger.  The
European Photon Imaging Camera (EPIC) CCD cameras were operated in the
Full Frame Mode. The total exposure time for the EPIC-pn camera was
14.0 ks. The Thin1 optical blocking filter was used for the EPIC-pn
camera, whereas the medium filter was used for both EPIC-MOS
cameras. The \emph{XMM-Newton} Observation Data Files (ODF) were
processed using \emph{XMM-Newton} Science Analysis Software (SAS
version 7.1.0) and the calibration files from the Calibration Access
Layer as on 14 Dec 2007. After excising periods of high-background,
the net exposure times of the EPIC-pn, MOS1, and MOS2 observations are
reduced to 3.6 ks, 10.8 ks, and 7.9 ks, respectively.

The \emph{XMM-Newton} EPIC observations of GRB\,080913 detected X-ray emission
from its afterglow (see Fig. 2) at EPIC-pn, MOS1, and MOS2 count rates of
0.0109$\pm$0.0021 cnts~s$^{-1}$, 0.0032$\pm$0.0008 cnts~s$^{-1}$, and
0.0019$\pm$0.0008 cnts~s$^{-1}$, respectively.  The net count number, $\sim$90
counts, is not sufficient to carry out a spectral fit.  Adopting an absorbed
power-law model of spectral index $\Gamma$=1.7 and column density
$N_{\rm H} =3.2 \times 10^{20}$~cm$^{-2}$, the EPIC spectral shapes and count
rates imply an absorbed X-ray flux of
3.8$\times$10$^{-14}$~ergs~cm$^{-2}$~s$^{-1}$ and an unabsorbed X-ray flux of
4.1$\times$10$^{-14}$~ergs~cm$^{-2}$~s$^{-1}$ in the energy band 0.3-10
keV. The X-ray luminosity in this same band at restframe is 5.7$\times$
10$^{45}$ erg s$^{-1}$ (assuming in XSPEC H$_0$ = 71 km/s/Mpc and
$\Omega_{\Lambda}$ = 0.73).

\section{Results and discussion} 
  \label{resultados} 
  
Following the discovery of the X-ray afterglow with {\it Swift}, we
detected a faint nIR afterglow (consistent with the position given by
Rossi et al. 2008) and the X-ray afterglow 4.5 days after with {\it
XMM-Newton}. We discuss in this section the classification and likely
progenitor of this burst.

\subsection{Spectral Flux Distribution of GRB~080913}
\label{sec:sed}

Using the model and methods described by J\'ohannesson et al. (2006)
we fitted the multiband observations of the afterglow to a fireball
model with energy injections.

\begin{figure}[h] 
\begin{center}
       \includegraphics[width=8.5cm,angle=0]{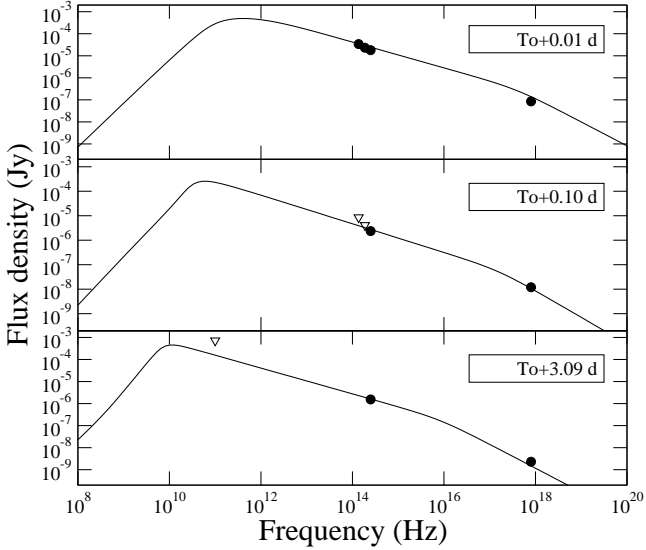}

\vspace{0.8cm}
      
         \caption{Spectral Flux Distribution of the afterglow from
radio to X-rays 0.01, 0.10, and 3.09 days after the burst in the
observer frame.  Filled circles are based on the nIR data from Greiner
et al. (2009) and the X-ray data from {\it Swift}/XRT. Triangles
represent upper limit: in the T$_{0}$+0.01 days plot they correspond
to nIR observation upper limits whereas in the T$_{0}$+3.09 days plot
it is the 3-$\sigma $ upper limit obtained at PdB (this paper).}
    \label{sed}
\end{center} 
\end{figure}

One injection is needed in order to account for the bump seen at $\sim
0.5$ days in the light curves.  Since no jet-break is seen in the
light curves up to around 10 days, we can only put a lower limit to
the collimation angle of the jet, $\theta_0$ $\gtrsim$ 3$^\circ$.
This is delayed from the definition of Sari et al. (1999) due to the
energy injection, the sideways expansion of the jet and the detailed
calculation of the equal arrival time surface.  Our preferred scenario
(giving the best fit) for a collimation of $3^\circ$ is an initial
energy release of $E_0$ = 4$\times$ 10$^{51}$ erg into a uniform
medium with density $n_0 = 2$ cm$^{-3}$.  This fit results in a
$\chi^2/$d.o.f = 170/50 where the high value is mainly caused by the
scatter in the X-ray light curve the model is unable to reproduce.  In
order to explain the bump at 0.5 days, an energy injection carrying
approximately $12 E_0$ (5$\times$ $10^{52}$ erg) is needed.  Due to
the lack of a turnover in the light curves at early time, we require
the initial Lorentz factor of the blast wave to be $\Gamma_0$
$\gtrsim$ 500.  The SFD is best fit with an electron index $p = 2.17$
and micro- physical parameters $\epsilon_i$ = 6$\times$$10^{-4}$ and
$\epsilon_B$ = 3$\times$10$^{-5}$.  Note that the parametrization of
the electron population has changed from Johannesson et al. (2006) and
we now follow the tradition of Panaitescu \& Kumar (2001).  The
minimum Lorentz factor of the electron distribution is now defined as
$\gamma_{\rm min} = \epsilon_i m_p/m_e (\Gamma -1)$.
Figure~\ref{sed} shows the radio to X-ray SFD predicted by our model
for 3 epochs together with observation data corrected for intrinsic
extinction.  The Galaxy extinction is negligible.

Please note the the above values for the best fit parameters are in
many cases highly dependent on the value chosen for $\theta_0$.  The
energy required for the blast wave goes as $\theta_0^2$ and lower
values of $\theta_0$ put an upper limit on $n_0$ from the jet break
requirements.  Additionally, there is a strong correlation between the
value of $n_0$ and $\epsilon_B$ because we do not have the required
data to constrain the SFD at lower frequencies.  Other parameters are
less sensitive to the value of $\theta_0$, especially $p$ that is well
constrained from the nIR SFD.

\vspace{3.5cm}

 \subsection{Comparison  with  high-redshift long GRB\,050904  and short  GRB\,060121}
           \label{comparison} 

\begin{figure}[t] 
\begin{center}
      \includegraphics[height=8.5cm,angle=-90]{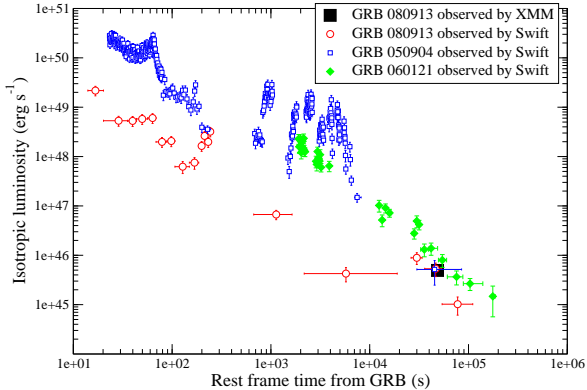}
      \caption{The K-corrected (rest frame) 
        0.3-10  keV   isotropic  luminosity  evolution  
        of GRB\,080913 (red) in comparison to the long-duration GRB\,050904 
        (blue, $z=6.3$)  and  the short-duration  GRB\,060121 (green, $z=4.6$). 
        The solid  square at $t-t_{GRB}=4.9\times  10^4 s$  represents the
        {\it XMM-Newton} observations carried out for GRB\,080913. As seen
        GRB\,080913 shows lower isotropic luminosities than the short
        GRB\,060121.}
    \label{lightcurve}
\end{center} 
\end{figure}

Given their highest redshifts, one could qualitatively compare the
properties of GRB\, 080913 ($z=6.7$), GRB\,060121 ($z=4.6$) and GRB\,
050904 ($z=6.295$).  The reason for choosing the latest cases is that
GRB\,060121 is a short duration GRB (T$_{90}$$\sim$2s), very likely at
high-redshift ($z=4.6$, de Ugarte Postigo et al. 2006), whereas GRB\,
050904 is a long GRB (T$_{90}$$\sim$31s ) at a comparable redshift
($z=6.295$; Haislip et al. 2006; Kawai et al. 2006).  We have
constructed the restframe isotropic 0.3-10 keV luminosity lightcurves
of the three GRBs, assuming a power law spectrum with a photon index
$\Gamma$.  The time evolution of $\Gamma$ was found by using linear
interpolation between the $\Gamma$ values determined from {\it Swift}
/XRT spectra on a logarithmic time scale.  This was done to get a
smooth K-correction with time. As seen in Fig.~\ref{lightcurve} the
0.3-10 keV decay of GRB\, 080913 is similar in character to those
exhibited by GRB\,060121 and GRB\, 050904.  However, its isotropic
luminosity at early stages is lower by a factor $\sim$30. Therefore,
based on the afterglow lightcurve it is not possible to
rule out any of the possible origins for the progenitor.

   \subsection{Implications for a short duration progenitor origin} 
         \label{implications_short} 

Because of the intrinsically short duration, hard spectrum and the
effectively zero spectral lags of GRB 080913, we consider in the
following the constraints on the typical progenitor systems invoked
for standard short-hard GRBs. We note that a late energy injection,
used to fit the afterglow light curve (Sect.~\ref{sec:sed}), does not
necessarily argues against a short GRB progenitor or, equivalently,
imply a long lived central engine, if the ejected material has a
distribution in initial Lorentz factors. Under the assumption that the
initial material powering the afterglow was ejected with a Lorentz
factor of $\sim 500$, matter ejected simultaneously with a Lorentz
factor of $\sim 40$ would refresh the afterglow at around
$0.5\,$days. The rather large energy injection of 10 times the initial
energy release requires, however, a significant increase in the
efficiency of the central engine that poses constraints on either
short or long lived central engines.

At a redshift $z$ = 6.7, the burst ocurred when the Universe was only
$\sim$ 0.8 Gyr old.  According to Yoshida et al. (2008), the first
stars formed $\sim 0.2$\,Gyr after the Big Bang.  Thus, if GRB\,080913
resulted from a merger of compact objects, the whole evolution of the
progenitor stars ($\tau_{\rm evol}$) and the merger time ($\tau_{\rm
mer}$) ha\-ppened within $\tau = \tau_{\rm evol} + \tau_{\rm mer}
\lesssim 0.6$\,Gyr. Such a small value of $\tau$ sets a lower limit to
the mass of the progenitor stars. Thus, neither of the two stars
forming the binary system could be less massive than $\sim 4
M_{\odot}$.  This only restricts the type of progenitor system if the
burst were a result of the merger of a white dwarf (WD) and a
BH.\footnote{WD+NS mergers possess much longer timescales and are more
likely progenitors of long GRBs (e.g., King, Olsson, \& Davies 2007)}

As the lower accretion timescales ($\sim 10$\,s) o\-ccur for the
higher-mass WD (Popham, Woosley \& Fryer 1999), the WD should have a
mass close to the Chandrasekar mass ($\sim 1.4 M_{\odot}$). For
NS+NS/BH binaries, the evolution of the progenitor stars is only a
small fraction of $\tau$ ($\tau_{\rm evol} \sim 10 - 20\,$\,Myr;
Belczynski et al. 2006). Thus, $\tau$ roughly equals $\tau_{\rm mer}$
in this case.

Since $\tau_{\rm mer}\sim 0.6\,$Gyr, if the GRB originated from a
double NS (DNS) merger, it could have proceeded through a classical
merger channel, for which $\tau_{\rm mer}\sim 0.1 - 15\,$Gyr
(Belczynski et al.  2006). Considering the large $E_{iso}$ of
GRB\,080913, it is more likely that it was hosted in a NS+BH merger
than in a DNS. The latter type of mergers have typically $\tau_{\rm
mer}^{NS+BH}\sim 1\,$Gyr, but with a non-negligible merger probability
for times $0.1 \rm{Gyr} \lesssim \tau_{\rm mer}^{NS+BH} \lesssim
1\,$Gyr.

According to Oechslin \& Janka (2006) the total equivalent isotropic
energy released in gamma-rays ($E_{iso}$) from a progenitor consisting
of a BH-accretion torus system, which boosts a neutrino driven
ultrarleativistic jet is
\begin{equation}
E_{iso} = f_1\, f_2 \,f_3\, f_4\, f_\Omega^{-1}M_{acc}c^2\, ,
\label{eq:eiso}
\end{equation}
where $M_{acc}$ and $c$ are the accreted mass and the light speed,
respectively, $f_1$, $f_2$, $f_3$, and $f_4$ are different efficiency
factors (see below), and $f_\Omega\simeq \theta_0^2/2$ is the jet
collimation factor. For GRB~080913, both $f_\Omega$ and $f_4$, the
fraction of the energy of ultrarelativistic jet matter, which can be
emitted in gamma-rays in course of dissipative processes that occur in
shocks, can be fixed from the light curve fit shown in
Sect.~\ref{sec:sed}.  A jet collimation factor
$f_\Omega=1.37\times10^{-3}$ results taking as jet half-opening angle
$\theta_0 = 3^\circ$ (Sect.~\ref{sec:sed}). This value is smaller
than, but consistent with $f_\Omega\sim 0.015 - 0.034$ obtained in
numerical models (Aloy, Janka \& M\"uller 2005), since $\theta_0$ is
only a lower bound in our case.  The factor $f_4$ is set by the ratio
$f_4=f_\Omega E_\gamma/E_{AG}=0.024$, where $E_{AG}$ is taken here
equal to the kinetic energy used to model initial afterglow in
Sect.~\ref{sec:sed}, i.e., $E_{AG}=E_0$. Such figure is also
consistent with estimates of the internal shock model, $f_4\lesssim
0.3$ (e.g., Mimica \& Aloy 2009 and references therein). For the
remaining factors in Eq.~(\ref{eq:eiso}), we find that the observed
energy in the prompt GRB phase needs the concurrence of a large
accretion disk mass ($M_{acc}\gtrsim 0.55 M_{\odot}$; Oechslin, Janka
\& Marek (2007) obtain $M_{acc}\lesssim 0.3 M_\odot$) and several {\it
  large} efficiency factors (probably so large that they rule out a
neutrino mediated energy extraction from the central engine). A large
conversion efficiency of the accreted mass into neutrino emission
$f_1\gtrsim 0.1$ (to be compared with a typical value $f_1 \gtrsim
0.05$; e.g. Setiawan et al. 2006; Lee et al. 2005), a conversion
efficiency of neutrinos and antineutrinos by annihilation to $e^\pm$
pairs $f_2\gtrsim 0.06$ (for reference, $f_2 \sim 0.001, \ldots ,
0.04$ is estimated by Ruffert \& Janka (1999), Setiawan et al. (2006),
or Birkl et al. (2007), and a large fraction of the $e^\pm$-photon
fireball energy which drives the ultra-relativistic outflow
$f_3\gtrsim 0.4$ (larger than $f_3\sim 0.1$ extracted from
simulations; Aloy, Janka \& M\"uller 2005). The combination of a large
$f_2$ and a large disk mass is supported by the steady models of Birkl
et al. (2007) with values of the dimensionless angular momentum of the
central BH $a\sim 0.4 - 0.5$.

We note that if $E_{AG}$ is taken to be the sum of the contributions
due to the initial kinetic energy ($E_0$) and of the late energy
injection ($12 E_0$; Sect.~\ref{sec:sed}), then $f_4 \sim 1.3\times
10^{-3}$, implying that all the remaining {\it free} efficiency
factors ($f_1$, $f_2$ and $f_3$) have to be, at least, a factor of 5
larger than the largest estimates of them, which suggests that a
neutrino driven outflow cannot account for the observed phenomenolgy.

We point out that the previous analysis is only sensitive to the value
of $\theta_0$ indirectly, if we seek to accommodate with our model the
values of $E_{AG}$. This is because $E_{AG} \simeq f_1 f_2 f_3
M_{acc}c^2$, but the fitted values of $E_{AG}$ depend on $\theta_0^2$
(Sect.~\ref{sec:sed}). Berger (2007) makes the hypothesis that
outflows of short-duration GRBs with the highest energies are strongly
collimated.  If the outflow is neutrino-driven, we point out that such
hypothesis seems consistent with the fact that the largest $E_{iso}$
are linked to BH-torus systems in which either the torus mass is large
or $f_1$ is large.  A large torus mass may arise in mergers between
compact objects of different masses (Oechslin \& Janka 2006; Shibata
\& Taniguchi 2008).  The conversion efficiency of the accreted mass
into neutrino emission tends to increase with increasing values of the
torus viscosity (Setiawan, Ruffert \& Janka, 2006).  Remarkably, a
large viscosity yields more vertically extended (inflated) accretion
tori, which may help to collimate the ultra-relativistic outflow in
narrow channels.

Given the large efficiencies needed to account for the large $E_{iso}$
of GRB\,080913 if the outflow were neutrino driven, it seems more
natural in this subclass of extremely energetic short GRBs to invoke
an energy extraction mechanism directly linked to the BH spin, e.g.,
the Blandford-Znajek (BZ) process (Blandford \& Znajek 1977). In this
case, an estimate of the total power produced by the central engine is
(Lee, Wijers \& Brown 2000), $P_{BZ}=1.7\times10^{50} a^2
(B/10^{15}\,\mbox{G})^2 (M_{bh} / M_\odot)^2 f(h)$, where
$f(h)=[(1+h^2)/h^2][(h+1/h) \arctan{(h)} - 1]$, $H = a
/(1+\sqrt{(1-a^2)}$, and $B$ is the magnetic field strength at the
event horizon of the BH with mass $M_{bh}$. We obtain the isotropic
equivalent energy released in $\gamma$-rays due to this process during
the intrinsic event duration ($T_{90}\simeq 1$\,s) as
$E_{iso}=P_{BZ}*T_{90}*f_3*f_4$. Note that in the previous estimate,
we use the same efficiency factors $f_3\sim 0.4$ and $f_4=0.024$ as in
the previous paragraphs to account for the facts that (1) only a
fraction of the released energy will be used to drive a
ultrarelativistic outflow, and that (2) the radiated energy in
$\gamma$-rays is much smaller than the kinetic energy of the outflow
$E_{AG}$. Hence, the observed energy can be reached if either the
value of the dimensionless angular momentum of the central BH is $a
\sim 1$, the magnetic field surrounding the BH is $B \gtrsim
10^{16}\,$G or the BH has a mass $M_{bh}\gtrsim 20 M_{\odot}$.  We
note that to form a $20\,M_\odot$ BH in a low-metallicity star (which
shall be the case at the redshift of GRB~080913), the initial mass of
such star shall be $\gtrsim 50\,M_\odot$ (Woosley, Heger \& Weaver
2002).

Since $P_{BZ}$ depends quadratically on both $B$ and $M_{bh}$, we may
estimate which is the range of variation of these two parameters such
that the resulting $E_{iso}$ complies with the energetics observed for
GRB~080913. Lower values of the BH mass, $M_{bh}\sim 3 M_\odot$,
closer to the typically considered ones in mergers of compact objects,
require extremely large values of the magnetic field strength ($B
\gtrsim 7\times 10^{16}\,$G). Such magnetic fields would probably
brake excesively the rotation of the stellar progenitor core, likely
inhibiting the formation of a maximally rotating Kerr BH (i.e.,
reducing the value of $a$ sensitively below 1). Hence, in account of
the large estimated mass of the BH, if a merger of compact objects
were the progenitor system of this GRB, a NS+BH merger is favoured,
since the typical mass of the BH resulting from a DNS merger is
$\lesssim 3M_\odot$.  On the other hand, if we consider values of the
magnetic field strengh smaller than the reference value of
$10^{16}\,$G, we find that even a factor of three smaller field yields
a BH mass $M_{bh}\sim 60\,M_\odot$. To form such a massive BH the
progenitor star should have a mass $\sim 140 M_\odot$, i.e., in the
limit of being stars which are disrupted at the end of their lives by
the pair instability without leaving any remnant BH (Woosley, Heger \&
Weaver 2002).  Thus, both the magnetic field strength and the BH mass
are restricted within a relatively narrow range, $3\times
10^{15}\,\rm{G} \lesssim B \lesssim 3\times 10^{16}\,$G, and
$7\,M_\odot \lesssim M_{bh} \lesssim 60\,M_\odot$,
respectively. Within the former range of values for the triad of
parameters $a$, $B$ and $M_{hb}$, the most likely ones are those
favouring the largest possible value of $a$. Because of the fact that
strong magnetic fields tend to slow down the rotation of the stellar
core, and because of the difficulty to build up magnetic fields in
excess of $\sim 10^{15}\,$G by the collapse of stellar cores (e.g.,
Obergaulinger, Aloy \& M\"uller 2006, Obergaulinger et al. 2006), even
considering the action of the magnetorotational instability (Akiyama
et al. 2003; Obergaulinger et al. 2009), values of $a\sim 1$ fit
better with the lowest values of $B$ in the aforementioned
range. Thereby, to reach the appropriate BZ power to fuel GRB~080913,
we favour BH masses in the upper end of the interval stated above.

As noted above, if the total energy of the afterglow is $E_{AG} \simeq
13 E_0$, the reduced value of $f_4$, yields even more stringent
constraints on the central engine, since $P_{BZ}$ has to be 13 times
larger, which needs of $a\sim 1$, $B \gtrsim 1.6 \times 10^{16}\,$G
and $M_{bh}\gtrsim 50 M_{\odot}$. However, we have to be cautious with
the inferences based on the values of the total kinetic energy in the
afterglow, since differently from $E_{iso}$ (directly measured),
$E_{AG}$ results from a model fit of the afterglow light curve, which
sensitively depends on the value of $\theta_0$ (Sect.~\ref{sec:sed}).

Finally, the large redshift of GRB080913 fits in theoretical models
where the rate of NS+NS/BH mergers follows either the star formation
rate or the star formation rate with delays smaller than $1\,$Gyr
(Janka et al. 2006).

\subsection{Implications for a long duration progenitor origin}
             \label{implications_long}

Recent studies (Zhang et al. 2009, Belczynski et al. 2009) have
suggested that GRB 080913 has a long-duration progenitor origin.

Zhang et al. (2009) present a new scheme for classifying bursts based
on criteria more closely related to the progenitor type to
differentiate physical origins, such as SN association, host galaxy
properties and the offset of the GRB location in the host galaxy. They
classify bursts into two main categories: Type I (with a compact
star-merger origin) and Type II (with a massive-star core collapse
origin). Under such scheme, GRB\,080913 and GRB\,090423 are identified
as Type II candidates based on (i) the geometrically-corrected
gamma-ray ($E_\gamma$) and kinetic ($E_K$) energies (with large
values), (ii) intrinsic afterglow luminosities (moderately bright),
(iii) the high density of the circumburst medium and, (iv) the
marginal compliance of the $E_p - E_{\gamma,iso}$ relation.  However,
because these bursts are intrinsically short, but still considered
more likely long, they suggested a possible ``specific Type I
scenario'' driven by the Blandford-Znajek mechanism of a BH-NS merger.

The analysis made in the previous section, regarding the properties of
the central BH in order to be able to deliver the sought $E_{iso}$ of
GRB 080913, is formally independent of the fact that the progenitor
system is a single massive star or a member of a binary. Thus, we
cannot exclude the possibility that the progenitor system is a massive
low-metallicity star that forms a collapsar-like engine (or a Type II
GRB according to the classification of Zhang et al. 2009) whose energy
is extracted by means of a BZ-mechanim. In such a case, the estimated
BH mass and magnetic field are the same as in
Sect.~\ref{implications_short}.

Belczynski et al. (2009) point out that based on the currently used
gamma-ray diagnostics (T$_{90}$, E$_{peak}$, hardness ratio) these
bursts would be considered as short had they occured at low
redshift. They argue that based on the average {\it Swift} detection
rates (accounting for selection effects), estimated for long GRBs to
exceed 10 times the rates for short GRBs, these bursts might belong to
the long class. At the redshifts of these GRBs, the calculated rates
are 1 yr$^{-1}$ and 0.1 yr$^{-1}$ per unit redshift, for long and
short GRBs, respectively.

\section{Conclusions} 
  \label{conclusiones} 
   
We report multiwavelength observations of the high $z$, potentially
short-duration gamma-ray burst \,GRB 080913 acquired between 20.7
hours and 16.8 days after its detection by \emph{Swift}. Although nIR
and X-ray afterglows were found, no mm afterglow was detected.  The
X-ray spectrum is consistent with negligible intrinsic absorption. We
have modeled the data with a collimated ($\theta_0$ $\gtrsim$ 3 deg)
blast wave with an energy injection at 0.5 days, requiring a total
energy release of more than $5\times 10^{52}$ erg.

At a redshift $z$ = 6.7, the burst ocurred when the Universe was only
$\sim$0.8 Gyr old.  If GRB\,080913 resulted from a merger of compact
objects, the whole evolution of the progenitor stars and the merger
time happened within $\lesssim$ 0.6 Gyr, following the star formation
rate with delays of $\lesssim$ 1 Gyr (Janka et al. 2006).  This
implies that neither of the two stars could be less massive than $\sim
4 M_{\odot}$. If the system of compact binaries were formed by a WD
and a BH, the short duration of the burst restricts the mass of the WD
to be close to the Chandrasekar mass. If the progenitor system of this
GRB is a merger of compact objects, a NS+BH system is favored over a
DNS mergers, having the BH a rather large mass ($M_{bh}\gtrsim
20\,M_\odot$) and being threaded by huge magnetic fields ($B\gtrsim
10^{16}\,$G). An electromagnetic energy extraction mechanism directly
linked to the BH spin, like the Blandford-Znajek process, may likely
operate in this subclass of GRBs. We note that only a factor of three
smaller value of $B$, would request $M_{bh}\gtrsim 60\,M_\odot$ to
explain the observed $E_{iso}$.  However, to form so massive BHs the
progenitor star shall have a mass $\sim 140\,M_\odot$, i.e., they are
in the limit of being stars whose final fate is to be disrupted by a
single pulse due to the pair instability, without leaving any remnant
BH (Woosley, Heger \& Weaver 2002), i.e., without the possibility of
producing a GRB engine. Considering the fact that magnetic fields in
excess of $\sim 10^{16}\,$G are difficult to build up by the collapse
of the core of the progenitor star, and that such fields may reduce
the angular momentum of the resulting BH, our preferred choice of
parameters is a combination of magnetic field strenght $B\gtrsim
\rm{few}\, \times 10^{15}\,$G and $M_{bh}\gtrsim 50\,M_\odot$. In
regard of these figures, we temptatively suggest that systems with
large BH masses (close to the limit set by the pair instability) and
magnetar-magnetic field strengths may constitute the {\it specific
  Type I scenario} invoking the BZ-mechanism in a NS+BH mergers
ocurring at high-$z$ considered by Zhang et al. (2009). However, we
cannot rule out the possibility that a single massive low-metallicity
star (Type II scenario), rather than a NS+BH merger, that yields a BH
with the aforementioned properties, and which fuels an
ultrarelativistic ejecta by means of a BZ-mechanism, constitutes the
central engine of GRB~080913.

It has been also demonstrated that the synergy between missions like
{\it Swift}, automated instruments and robotic observatories,
facilitate further study of the population of high-z GRBs, and help to
assess whether a significant fraction of short-duration GRBs coexist
at these high redshifts.

\begin{acknowledgements}

We thank the generous allocation of observing time by different Time
Allocation Committees.  This work is partially based on observations
collected at the Centro Astron\'omico Hispano Alem\'an (CAHA) at Calar
Alto, operated jointly by the Max-Planck Institut f\"ur Astronomie and
the Instituto de Astrof\'{\i}sica de Andaluc\'{\i}a (CSIC). IRAM is an
international institute funded by the Centre National de la Recherche
Scientifique (France), the Max Planck Gesellschaft (Germany) and the
Instituto Geogr\'afico Nacional (Spain). This work made use of data
supplied by the UK {\it Swift} Science Data Centre at the University
of Leicester and of data products from the Two Micron All Sky Survey
(2MASS), 2MASS, which is a joint project of the Univ. of Massachusetts
and the IR Processing and Analysis Center/CalTech, funded by NASA and
NSF.  We thank the assistance of D. Crist\'obal Hornillos. DPR
acknowledges support from the ``Jos\'e Castillejo'' program. AdUP
acknowledges support from an ESO fellowship. MAA is a Ram\'on y Cajal
Fellow.  JO and KP acknowledge the support of the STFC. IH and VP
acknowledge support from OTKA grants T48870 and K77795. This research
has also been partially supported by the Spanish MICINN under the
programmes AYA2007-63677, AYA2008-03467/ESP, AYA2007-67626-C03-01 and
CSD2007-00050.
 
\end{acknowledgements}

\end{document}